\documentclass[aps,twocolumn,graphics,floatfix,tightenlines]{revtex4-2}
\usepackage{graphics}
\usepackage{epstopdf}
\usepackage{epsfig}
\usepackage{graphicx}
\usepackage{epsf,epic}
\usepackage{color}
\usepackage{subfig}
\usepackage{amsmath}
\usepackage{booktabs}
\usepackage{multirow}
\usepackage{physics}
\usepackage{hyperref}
\usepackage{amsfonts}
\usepackage{wrapfig}
\usepackage{breqn}
\usepackage{pstricks}
\usepackage[utf8x]{inputenc}
\usepackage{multirow}
\usepackage{fancyref}
\usepackage{pst-node}
\usepackage{soul}
\usepackage{float}
\usepackage{bm}
\usepackage{dcolumn}
\newcommand{\etal}{\textit{et al.\ }}
\newcommand{\ie}{\textit{i.e.\ }}

\makeatletter
\usepackage{etoolbox} 
\appto{\appendix}{%
	\@ifstar{\def\theequation@prefix{A.}}%
	{}%
}
\preto\maketitle{%
  \begingroup\lccode`~=`,
  \lowercase{\endgroup
  \let\saved@breqn@active@comma~
  \let~}\active@comma 
}
\appto\maketitle{%
  \begingroup\lccode`~=`,
  \lowercase{\endgroup
  \let~}\saved@breqn@active@comma 
}
\makeatother
\begin{document}
\title{Optical absorption and luminescence of $\alpha$-LiV$_2$O$_5$ from the Bethe Salpeter Equation}

\author{Claudio Garcia}
\author{Walter R L. Lambrecht}
\email{walter.lambrecht@case.edu}
\affiliation{Department of Physics, Case Western Reserve University, 10900 Euclid Avenue, Cleveland, OH-44106-7079, USA}

\begin{abstract}
  $\alpha$-Li$_x$V$_2$O$_5$ is obtained by intercalating Li between the layers of V$_2$O$_5$. The partial filling of the split-off conduction band  by electron donation from Li leads to significant changes in optical properties.  Here we study the electronic band structure of $\alpha$-LiV$_2$O$_5$ using quasiparticle self-consistent (QS)  $GW$ calculations and the optical dielectric function by means of the Bethe-Salpeter Equation (BSE). The half-filling of the narrow split-off band leads to a spin-splitting and formation of magnetic moments of 0.5 $\mu_B$ per V which order antiferromagnetically along the chain  or $b$-direction. 
  The imaginary part of the dielectric function shows a very strong optical absorption band near 2 eV for polarization along the $a$-direction  already in the independent particle approximation but red shifted in the BSE. It is about ten times stronger in intensity than the lowest exciton peaks in pure V$_2$O$_5$ or than the $b$-direction. This absorptions stems from a localized transition between the occupied V-$d_{xy}$ derived band, which is odd with respect to the $a$-mirror plane to the higher lying empty band formed from the same  V-$d_{xy}$
  orbitals but even with respect to that mirror-plane, which explains its polarization and large oscillator strength. It is found both for the ferromagnetic and antiferromagnetic arrangement of the spins along the $b$-direction.
  We relate our main finding  to a recent experimental study of
  cathodoluminescence (CL) in which a suppression of the lowest CL peak was observed  upon addition of Li. Based on our and literature results, we  present  a different interpretation of the CL
  peaks from that study, which was based on the orbital character in a band-to-band picture.  The lowest CL peak near 1.8 eV, which lies well below the indirect absorption onset of V$_2$O$_5$ is  proposed to be related to recombination of a self-trapped electron polaron, resulting from oxygen vacancies,  with a hole at the valence band maximum  and is suppressed in LiV$_2$O$_5$ by the strong self--absorption from the Li induced  occupied band to the higher empty bands at about the same energy.
\end{abstract}
\maketitle
\section{Introduction}
\begin{figure}
(a)\includegraphics[width=7cm]{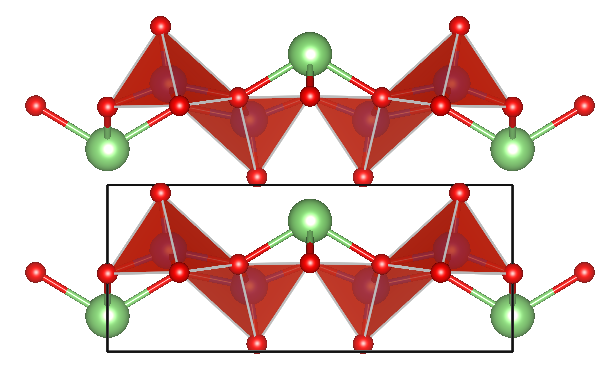}
(b)\includegraphics[width=7cm]{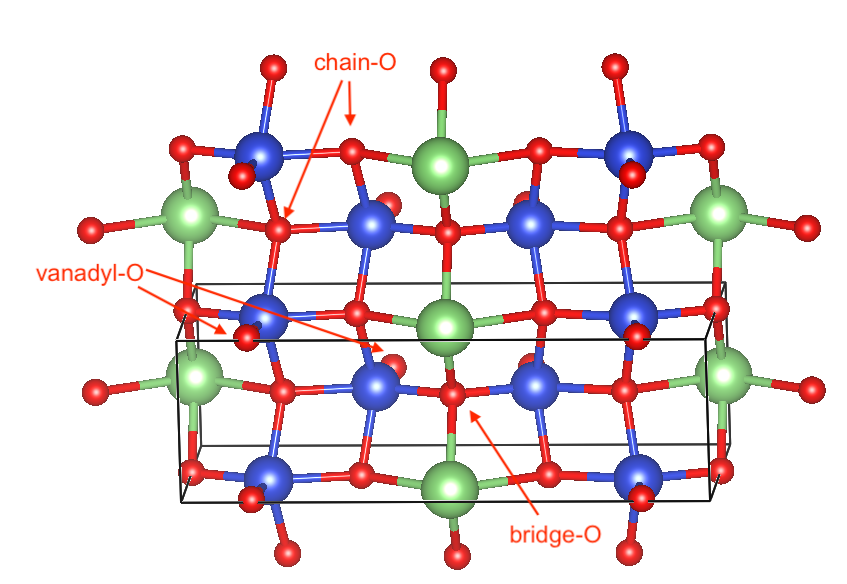}
\caption{Structure of $\alpha$-LiV$_2$O$_5$: (a) side view with V-surrounding polyhedra and (b) top view (ball and stick): green spheres Li, blue spheres V, red spheres O.\label{figstrucliv2o5}}
\end{figure}
Vanadium pentoxide or V$_2$O$_5$ is a layered oxide with strongly anisotropic properties in the layers thanks to the 1D zigzag chains of edge connected alternating up and down pointing square pyramids surrounding each V.\cite{Sucharitakul17}
Its properties can be significantly tuned by intercalating with Li, Na, Mg
and other ions. These atoms are located in between the  van der Waals bonded layers in the large open channels between the double zigzag chains, which are connected via a bridge oxygen. See Fig. \ref{figstrucliv2o5} for the crystal structure.
These ions can thus easily diffuse in and out of the structure and this provides a way to tune the properties of the material
forming so-called V$_2$O$_5$ bronzes. This makes Li$_x$V$_2$O$_5$ attractive as a Li cathode material, since up to a certain concentration, the  Li insertion and removal is reversible without significant structural changes. The study of V$_2$O$_5$ as a host for Li  cathode battery material goes back to the work
of the chemistry Nobel Prize winner Wittingham \cite{Whittingham_1976,Zavalij_1999} and has recently been re-activated by various authors.\cite{Walker_2020,Jarry_2020,Levy_2023,Warecki_2024,Gillette_2016,Rubloff_2013,Yue_2017,Rocquefelte_2003}
Optical characterization can play an important role in elucidating the electrochemical
insertion and extraction of Li.

Walker \etal \cite{Walker_2020} used depth resolved Cathodoluminescence
(CL) to study the evolution of the CL signal as function of Li uptake and the accompanying changes in
the structure of Li$_x$V$_2$O$_5$ were also studied by Raman spectroscopy by Jarry \etal\cite{Jarry_2020}.
The modifications of the optical properties observed in this work can be utilized for electrochromic applications as they modify the color of the crystals.
Walker \etal\cite{Walker_2020} observed a decrease in the  lowest energy CL peak (1.8-1.9 eV) upon Li uptake
which they analyzed in terms of local density approximation (LDA) band structure calculations by Eyert and H\"ock \cite{Eyert98} and recent X-ray edge 
spectroscopy studies of Li$_x$V$_2$O$_5$ \cite{DeJesus_2016,Horrocks_2016}.
However, it has recently become clear that the optical properties of V$_2$O$_5$ are not well
described by a simple interband transition picture  based on band structure theory. Quasiparticle self-consistent $GW$ calculations by Bhandari \etal\cite{Bhandari15} showed that the quasiparticle gap
is much larger ($\sim$4.4 eV) than the optical absorption edge of 2.3 eV extracted from Tauc plots by Kenny and Kannewurf\cite{Kenny66}.
Recently, this discrepancy was resolved by pointing out the large excitonic binding energies in this material\cite{Gorelov22,Garcia_2024}. The interpretation of the optical properties of Li$_x$V$_2$O$_5$ thus
requires  revision in terms of our current understanding of the excitonic properties of V$_2$O$_5$.
Polaronic effects are also known to play an important role in V$_2$O$_5$. Since early work on the transport it is known that the electronic transport is essentially of the variable range hopping type associated with small polarons.\cite{Ioffe_1970}
The strong electron-phonon coupling in V$_2$O$_5$ was shown to lead to self-trapped polarons by Scanlon \etal \cite{Scanlon08} and later further studied by Ngamwongwan \etal \cite{Ngamwongwan_2021} and Watthaisong \etal\cite{Watthaisong_2019}.

In this paper we calculate the electronic structure of LiV$_2$O$_5$ using the QS$GW$ method\cite{Kotani07} and evaluate the optical absorption properties using the  recent implementation  \cite{Cunningham18,Cunningham23} of the Bethe-Salpeter Equation approach based on the
linearized-muffin-tin-orbital (LMTO) basis set \cite{questaalpaper}.

First, we note that  for one Li per V$_2$O$_5$ unit the lowest split-off conduction band becomes exactly half-filled and this leads to the formation of a magnetic moment and a
spin-polarized band structure. In analogy with NaV$_2$O$_5$, we find  the spins to be ordered
antiferromagnetically along the chains\cite{Bhandari_2015_Na}.  While the magnetic properties  in LiV$_2$O$_5$ deserve a 
further study we here adopt this same spin arrangement and study the optical properties in both
ferromagnetic (FM) and antiferromagnetic (AFM) LiV$_2$O$_5$.  It is known from various studies of the lithiation \cite{Yue_2017,Rocquefelte_2003,Delmas_1994}
that upon higher concentration of Li, a phase transition occurs to
first the $\gamma$-phase of LiV$_2$O$_5$, which can already accommodate more than one Li per V$_2$O$_5$ unit. For
Li$_2$V$_2$O$_5$, the split-off band would become completely filled and magnetic moment formation is then no longer evident.
At even higher concentration Li$_3$V$_2$O$_5$ a non-layered (disordered rocksalt-type)  $\omega$-structure  has been obtained \cite{Rocquefelte_2003,Yue_2017,Delmas_1994,Galy_1992} In the present paper we only consider the regime where the structure stays close to that of $\alpha$-V$_2$O$_5$,
\ie $x\le1$ although
even there small modifications occur and have been labeled as the $\epsilon$- and $\delta$-phases.

Our primary goal is to provide an alternative interpretation for the CL key observation of a reduction of the lowest
CL peak at about 1.8 eV upon the initial lithiation.
\section{Computational Methods}
The band structures  are calculated using the QS$GW$ method \cite{MvSQSGWprl}, for which details
of the implementation are given in \cite{Kotani07}. This is a variant of Hedin's $GW$ method in which
$G$  is the one-electron Green's function and $W$ the screened Coulomb interaction \cite{Hedin65,Hedin69} and determine the dynamic self-energy $\Sigma=iGW$.
In the quasiparticle self-consistent approach, a non-local but energy independent exchange correlation potential
$\tilde{\Sigma}=\frac{1}{2}\sum_{ij}|\psi_i\rangle Re\left\{\Sigma_{ij}(\epsilon_i)+\Sigma_{ij}(\epsilon_j)\right\}\langle\psi_j|$
is extracted from the energy-dependent self-energy $\Sigma(\omega)$, with $|\psi_i\rangle$ the eigenstates of the non-interacting Hamiltonian $H_0$ and $Re$ meaning the Hermitian part. This allows to iteratively update and thereby optimize the $H_0$ starting from a density functional theory (DFT) functional such as LDA or generalized gradient approximation (GGA) and makes the final results independent of
the starting exchange-correlation functional choice. It focuses on the quasiparticle energies rather than the full self-energy or  interacting one-electron Green's function and this approach was further justified by Ismail-Beigi\cite{Ismail-Beigi_2017}. 

The
{\sc Questaal} code  implementation of this method used here uses a muffin-tin-orbital basis set and is described in \cite{questaalpaper}.
For the two-point quantities like the bare $v$ and screened $W=\varepsilon^{-1}v=(1-vP)^{-1}v$ Coulomb potentials and the polarization  $P$ and inverse dielectric
function $\varepsilon^{-1}$ it uses a mixed-product basis set, including plane waves restricted to the interstitial region and
products of partial waves times spherical harmonics in the muffin-tin spheres. This idea  originates from \cite{Aryasetiawan94}
and provides an efficient way to describe the screening at short range which does not require high energy partial waves or high energy
conduction bands. More recently, the Bethe Salpeter Equation (BSE) method \cite{Onida02}  which uses the two-particle Green's function to study the optical properties including local field and electron-hole interaction effects, was also implemented in the {\sc Questaal} code package as described in \cite{Cunningham18,Cunningham23}. In these papers, the BSE method is also used to include electron-hole interactions in the screening of $W({\bf q})$ by including ladder-diagrams and which is then called QS$G\hat{W}$. 
This reduces the QS$GW$ gap slightly but also decreases the exciton binding energies correspondingly.
Unlike our previous study
on V$_2$O$_5$ which used the QS$G\hat{W}$ method, we here use the original QS$GW$ method with RPA screening for $W$ because the  QS$G\hat{W}$ method becomes rather expensive for the large 64 atom cells needed for the antiferromagnetic structure.   While this increases the gaps (indirect, lowest direct and direct at $\Gamma$) of V$_2$O$_5$ by about 0.6 eV compared to the more accurate $\hat W$ approach, the exciton
binding energies are then also larger and the  energy of the lowest optical excitations are only overestimated by  $\sim$0.4 eV.
We are here focusing on qualitative changes due to Li and  avoided the more time consuming  BSE evaluation of screened $\hat W$.


\begin{figure*}
	\begin{minipage}{6.1cm}
		\includegraphics[width=5.7cm]{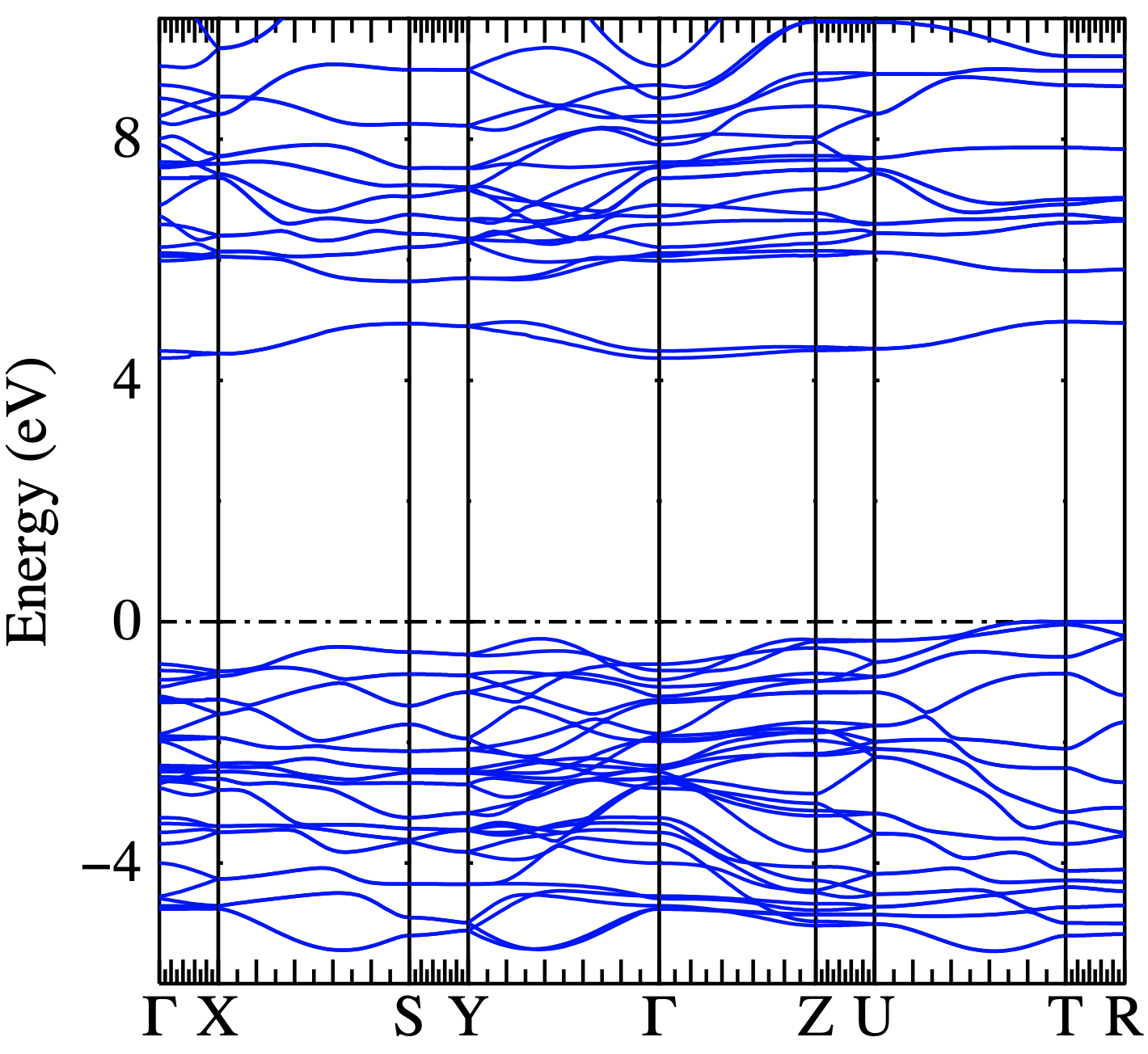}
		\vspace{0.1em}\\
		\hspace{-12.0em}(a)
	\end{minipage}\hskip -0.4cm
	\begin{minipage}{6.1cm}
		\includegraphics[width=5.7cm]{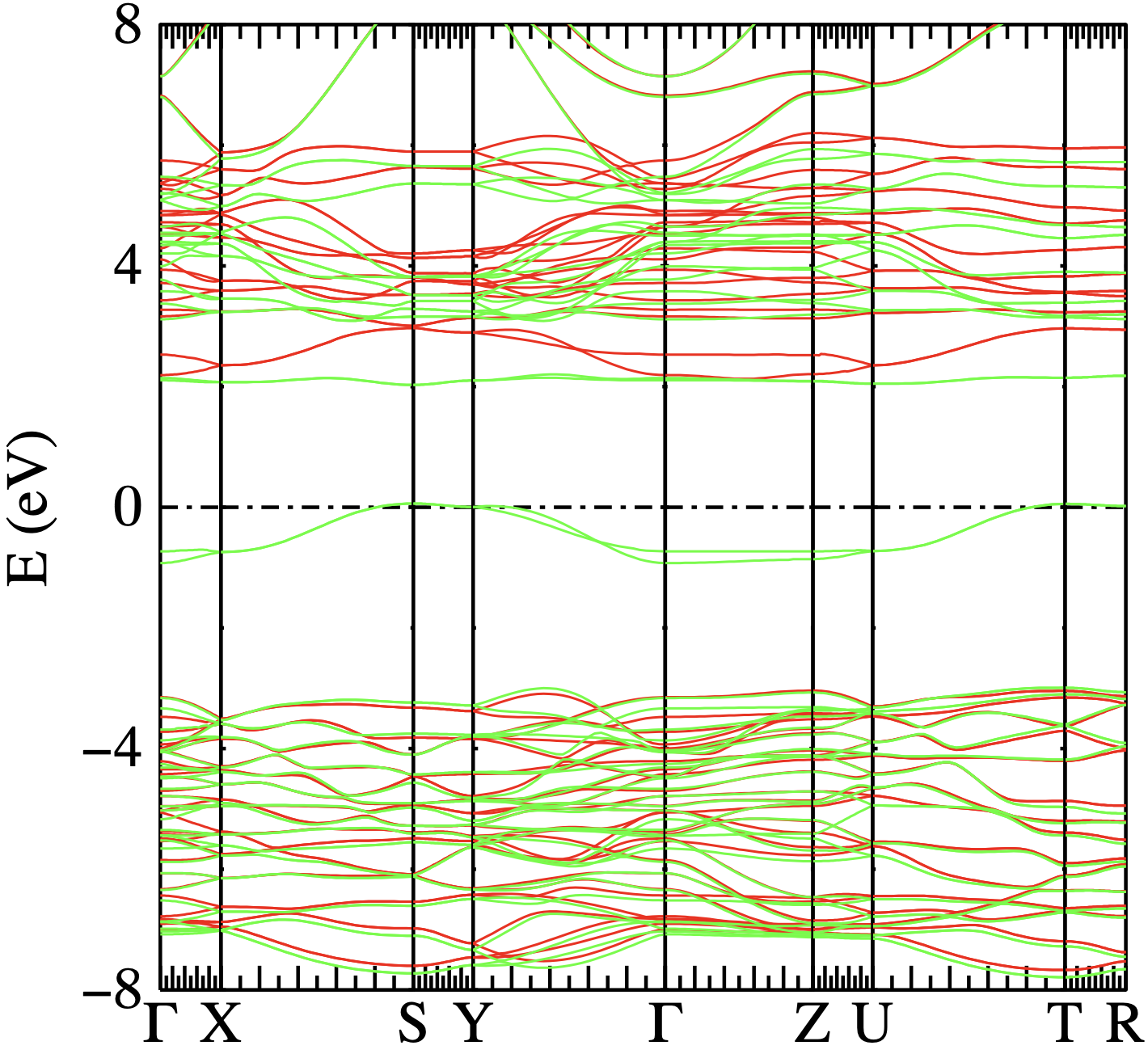}
		\vspace{0.1em}\\
		\hspace{-12.0em}(b)
	\end{minipage}\hskip -0.4cm
	\begin{minipage}{6.1cm}
		\includegraphics[width=5.7cm]{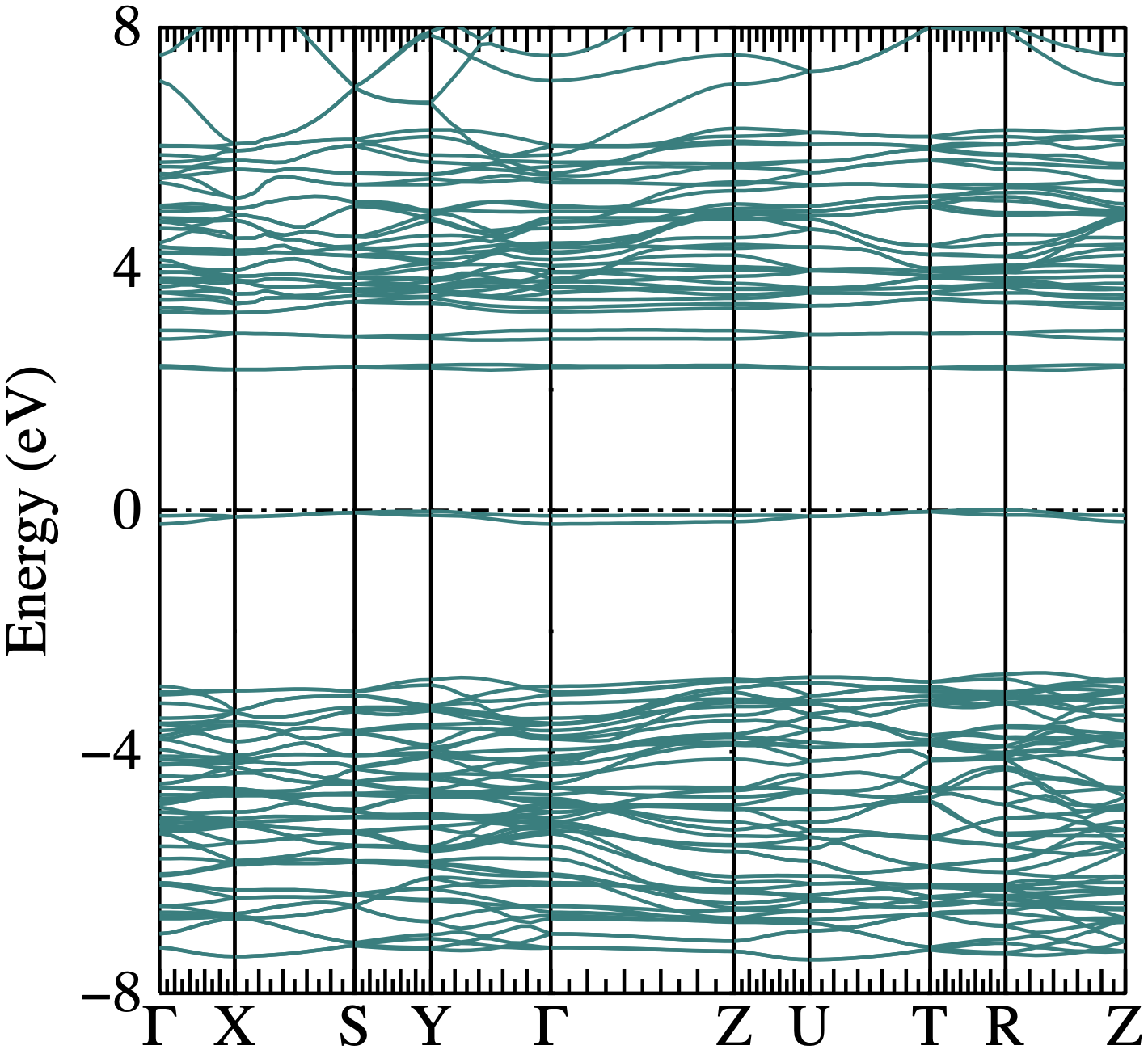}
		\vspace{0.1em}\\
		\hspace{-12.0em}(c)
	\end{minipage}
	\caption{QS$GW$ band structures of (a) V$_2$O$_5$, (b) FM LiV$_2$O$_5$, (c) AFM LiV$_2$O$_5$;  (a) blue non-spin-polarized, (b) green majority, red minority spin; (c) blue-green up and down spin superposed \label{figfmbnds}}
\end{figure*}

Details of the present calculation are similar to our recent study of optical properties in V$_2$O$_5$ \cite{Garcia_2024}. The convergence parameters were carefully optimized. We use a double set of smoothed Hankel function envelope functions with optimized decay lengths $\kappa$ and
associated smoothing radii with angular momenta $spdfspd$ for  V and O and $spdsp$ for  Li and include V-$3p$ semi-core states as local orbitals. We use a well converged {\bf k}-point mesh for the QS$GW$ and a cut-off of the self-energy of 2.13 Rydberg above which it is approximated by an average diagonal
value. The product basis set uses a default angular momentum cut-off of $l=4$ and a maximum reciprocal lattice vector for the interstitial plane wave part (the GCUTX parameter) of 2.6 Ry$^{1/2}$, while for the muffin-tin orbitals in the interstitial region in the $GW$ calculation, a cut-off (named GCUTB) of 3.2  Ry$^{1/2}$ is used.
Because of the  about 3 times larger unit cell $a$ lattice constant than  the $b$ and  $c$ lattice constants in the V$_2$O$_5$  cell  and  the small dispersion of the bands in the $c$  direction perpendicular to the layers, we use a smaller number of divisions in the reciprocal $a$ direction. Comparing a $1\times3\times3$ and a $1\times5\times5$ mesh we find that the gap in FM LiV$_2$O$_5$ case differs by less than 0.1 eV. The results shown in the figure correspond to the $1\times5\times5$ mesh which is adequately converged. 
  Note that a finer mesh such as $2\times 6\times 6$ is used for the DFT self-consistency and the self-energy is interpolated to this mesh in the course of the QS$GW$ iterations. A similar procedure is used to obtain  the bands along the symmetry lines by Fourier transforming to a real space representation and back to the {\bf k}-points along the symmetry lines. For the AFM cell which is doubled in the $b$ direction, our final calculations were done with a $1\times2\times3$ mesh. We start the QS$GW$ calculations with a GGA$+U$ calculation, using the Perdew-Burke-Ernzerhof (PBE)
\cite{PBE} functional
with a Hubbard $U$ of 0.1 Rydberg on the V-$d$ orbitals to initiate a magnetic moment, but which is then switched off as the $GW$ self-energy takes over the role of the Hubbard $U$. The results thus do not depend on this initial choice, which is only used to start from a
band structure with a spin-splitting due to the half-filling of the narrow split-off conduction band. For the BSE calculations we use a similar {\bf k}-mesh. While
BSE  for Wannier type excitons typically requires a fine mesh near the band extrema, this is not the case here because the band edges are very flat and the excitons are quite localized Frenkel type excitons.

\section{Results}
We start by comparing the QS$GW$  band structure results for pure V$_2$O$_5$ with those of  FM LiV$_2$O$_5$ shown in Fig. \ref{figfmbnds}a-b. The atomic orbital character of the bands in LiV$_2$O$_5$ is similar to that in V$_2$O$_5$ as discussed in \cite{Bhandari15} with the Li levels high in the conduction band and merely donating their electron to the lowest V-$3d$ bands. The manifold of occupied bands shown here consists of the O-$2p$ derived bands, and the conduction bands are mostly V-$3d$ derived with higher antibonding O-$2p$ character as we go higher in energy. 
As already announced in the introduction, we can see that the narrow split-off conduction band at about 4 eV above the VBM, splits in spin-up and spin down bands, the lower of which is then occupied by the extra electron donated by the Li. The exchange splitting of this band is about 2.9 eV. This is the splitting between the red and green bands in Fig.\ref{figfmbnds}b. The splitting between the lowest empty and the highest occupied majority spin bands  is about 2.8 eV at $\Gamma$ and 2.1 at $Y$, where it is lowest. This splitting is relevant to the optical transitions between equal spin bands.  In the $G_0W_0$ calculation, this gap is about 1.5 eV so, the self-consistency of QS$GW$ is important.  In pure V$_2$O$_5$ at $\Gamma$, the split-off band lies about 1.5 eV below the bottom of the main set of conduction bands while this splitting is lower
at $Y$, where it is about 0.8 eV in V$_2$O$_5$. So, this makes sense: at its maximum near $Y$, the split-off band in V$_2$O$_5$ is about 0.8 eV below the main CBM and the spin-splitting induced in LiV$_2$O$_5$ then places the majority spin at half the spin splitting (1.4 eV) below this point, so at about 2.2 eV below the main CBM of majority spin. 

We may also notice that the highest valence band which in pure V$_2$O$_5$ shows a significant dispersion along $\Gamma Z$
becomes flatter in LiV$_2$O$_5$. This indicates that the insertion of Li disrupts the interlayer hopping of the O-$2p$ like states.
The split-off band of majority spin has a band dispersion mostly along the $\Gamma Y$ direction, which is the $b$-direction along the chains
and has a band width of about 0.8 eV. In FM LiV$_2$O$_5$ at $\Gamma$ the bottom of the split-off band lies about 2.2 eV above the
rather flat VBM of the
O-$2p$ bands. The gap at $\Gamma$ from the O-$2p$ like VBM to the majority spin CBM at $\Gamma$ is 5.26 eV, which is slightly larger than
in pure V$_2$O$_5$ where it is 5.07 eV.
In a simple band-to-band transition picture, we may now expect new optical transitions from the occupied majority spin band to 
the lowest empty band of the same spin, at about 2 eV.
\begin{figure*}
  \includegraphics[width=12cm]{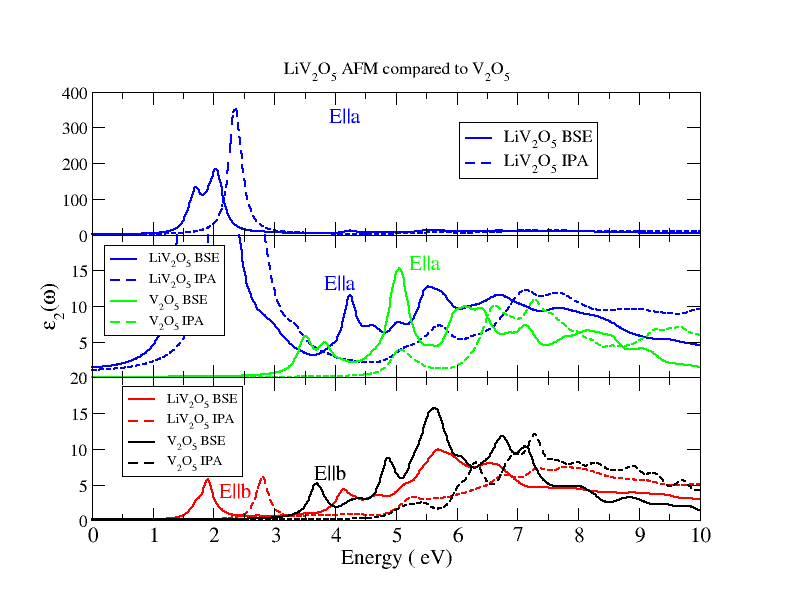}
  \caption{Imaginary part of the dielectric function for AFM LiV$_2$O$_5$ compared with V$_2$O$_5$ for ${\bf E}\parallel{\bf a}$ in the top two panels and ${\bf E}\parallel{\bf b}$ in the lower panel.  The second panel uses a different vertical scale to emphasize the higher energy range and allow for comparison with V$_2$O$_5$. Solid lines represent BSE and dashed lines IPA results. Blue and red correspond to LiV$_2$O$_5$ and black and green to V$_2$O$_5$. \label{figopticsafm}}
\end{figure*}

Next, we can see that in AFM LiV$_2$O$_5$, the  situation is essentially the same but now the spin-up and spin down bands are exactly superposed. Note that the spin up  bands on one V-O-V rung become the spin-down bands on the  next rung along the $b$-direction and vice versa. Because of the doubling of the cell along $b$, the bands are folded in two along $\Gamma Y$.  We may note that the highest occupied V-$d_{xy}$ derived bands become almost dispersion-less but the lowest gap is similar.
The band gap is 2.337 eV in QS$GW$, 1.846 eV in $G_0W_0$ and 0.728 eV in the initial GGA+U calculation. 

In terms of magnetic properties, we note that the net magnetic moment of the FM cell is 2 $\mu_B$. Since there are four V per cell,
this means effectively 0.5 $\mu_B$ per V. Indeed, because there are two formula units per unit cell, there are two
bands of majority spin that become exactly filled while their spin-down counterpart stays empty.
It indicates that we rather should consider the magnetic moment as being associated
with a single V-O-V rung and equally spread over the two V in that rung. This makes sense as the lower split-off band is essentially
a molecular type state formed by  V-$d_{xy}$ orbitals  which is odd versus the mirror plane perpendicular to the $a$-crystallographic axis. As such, it cannot interact with the O-$p_y$ states of the bridge oxygen between them, while the higher conduction band at the bottom
of the main conduction band manifold is an even combination of the same orbitals which does feel the antibonding $\pi$-type interaction
with the O-$p_y$ of the bridge. The O-$p_x$ and $p_z$ are orthogonal to the $xy$ orbitals of the two V along the $a=x$ direction
and do not come into play. The AFM state is found to have lower energy than the FM state and therefore we focus in the remainder on the AFM configuration.   Linear response function calculations indicate that the coupling between spins is antiferromagnetic along the $b$-direction, and this agrees with known experimental data in the related NaV$_2$O$_5$ and our findings here for LiV$_2$O$_5$.
We reserve a full discussion of the magnetic properties, such as the exchange interactions, spin waves and N\'eel temperature for  a separate paper.  We assume that even in the paramagnetic case, local magnetic moments form but  are disordered and the
electronic structure locally is similar to the one we here obtain for the FM case.

\begin{figure}
\includegraphics[width=8.6cm]{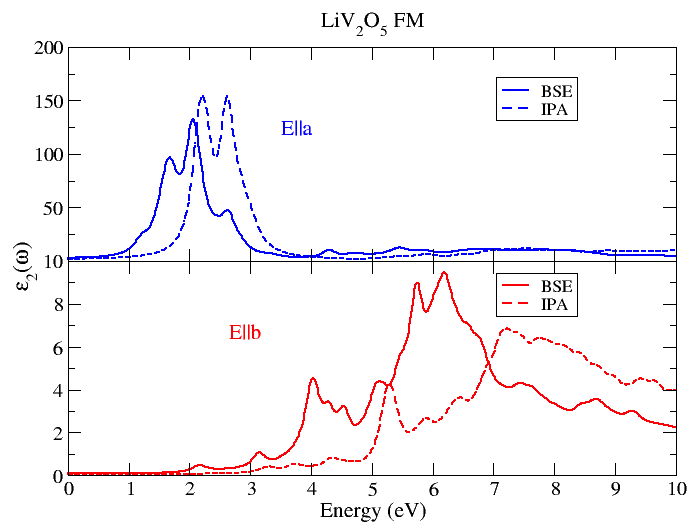}
  \caption{Imaginary part of the dielectric function for FM LiV$_2$O$_5$ \label{figopticsfm}}
\end{figure}
\begin{figure}
   \includegraphics[width=8.6cm]{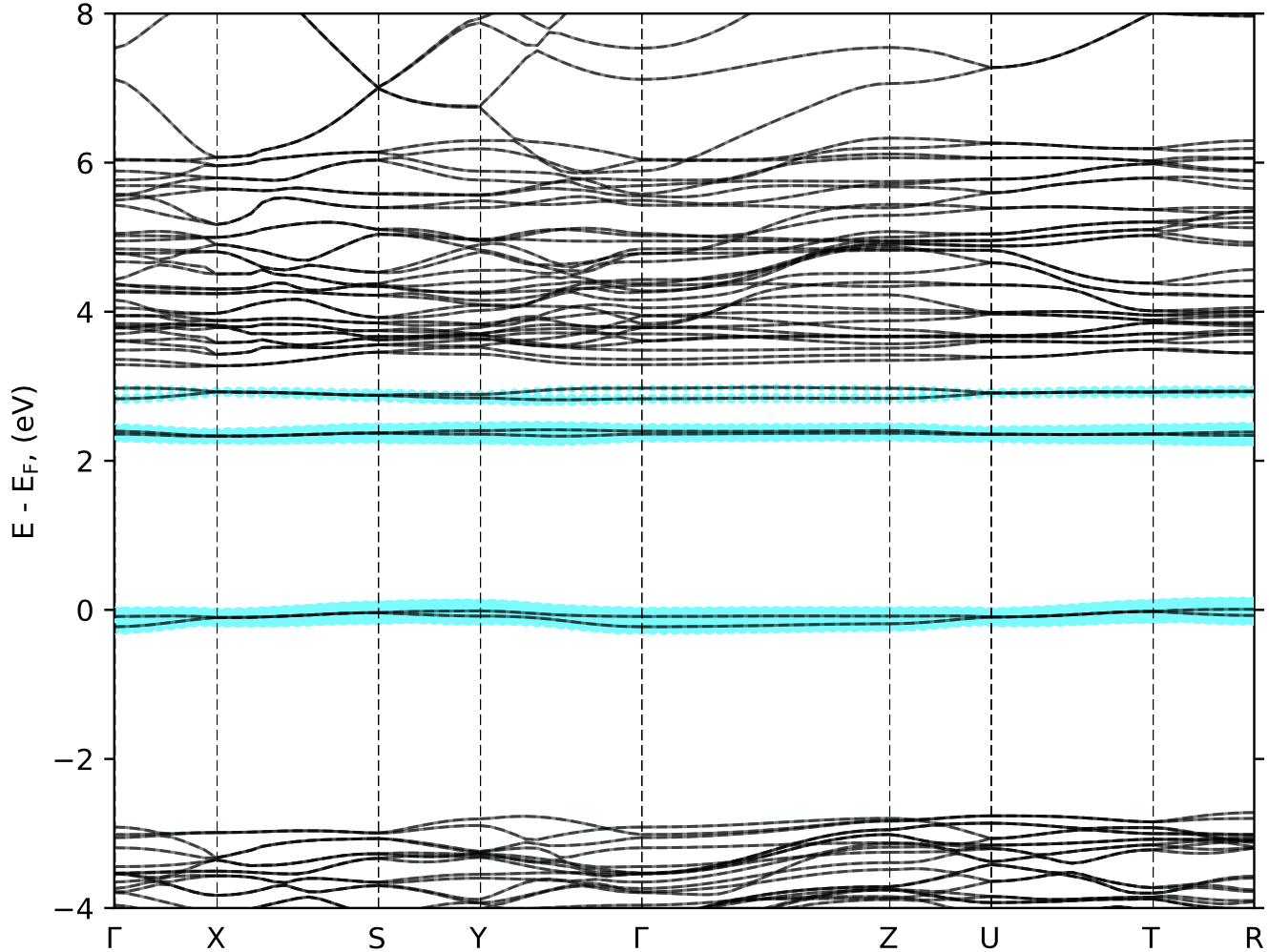}
 \caption{Band weights contributions to the peak in $\varepsilon_2(\omega)$
   in the range 1.7-1.85 eV. The width of the colored bands indicates the contribution
   $W_{c{\bf k}}=\sum_v  |A_{vc{\bf k}}^\lambda|^2$ for a given conduction  band $c{\bf k}$ of the two particle states in the energy range $E_{min}\le E^\lambda \le E_{max}$ covering the first peak  and with $A_{vc{\bf k}}^\lambda$ the eigenvectors of the BSE two-particle Hamiltonian which mixes vertical transitions at different {\bf k}-points. Similarly for occupied  bands,  $W_{v{\bf k}}=\sum_c  |A_{vc{\bf k}}^\lambda|^2$ gives the contribution of all conduction bands to transitions from a given valence band $v{\bf k}$.  
   \label{figbndwt}}
 \end{figure}

\begin{figure}
  \includegraphics[width=8.5cm]{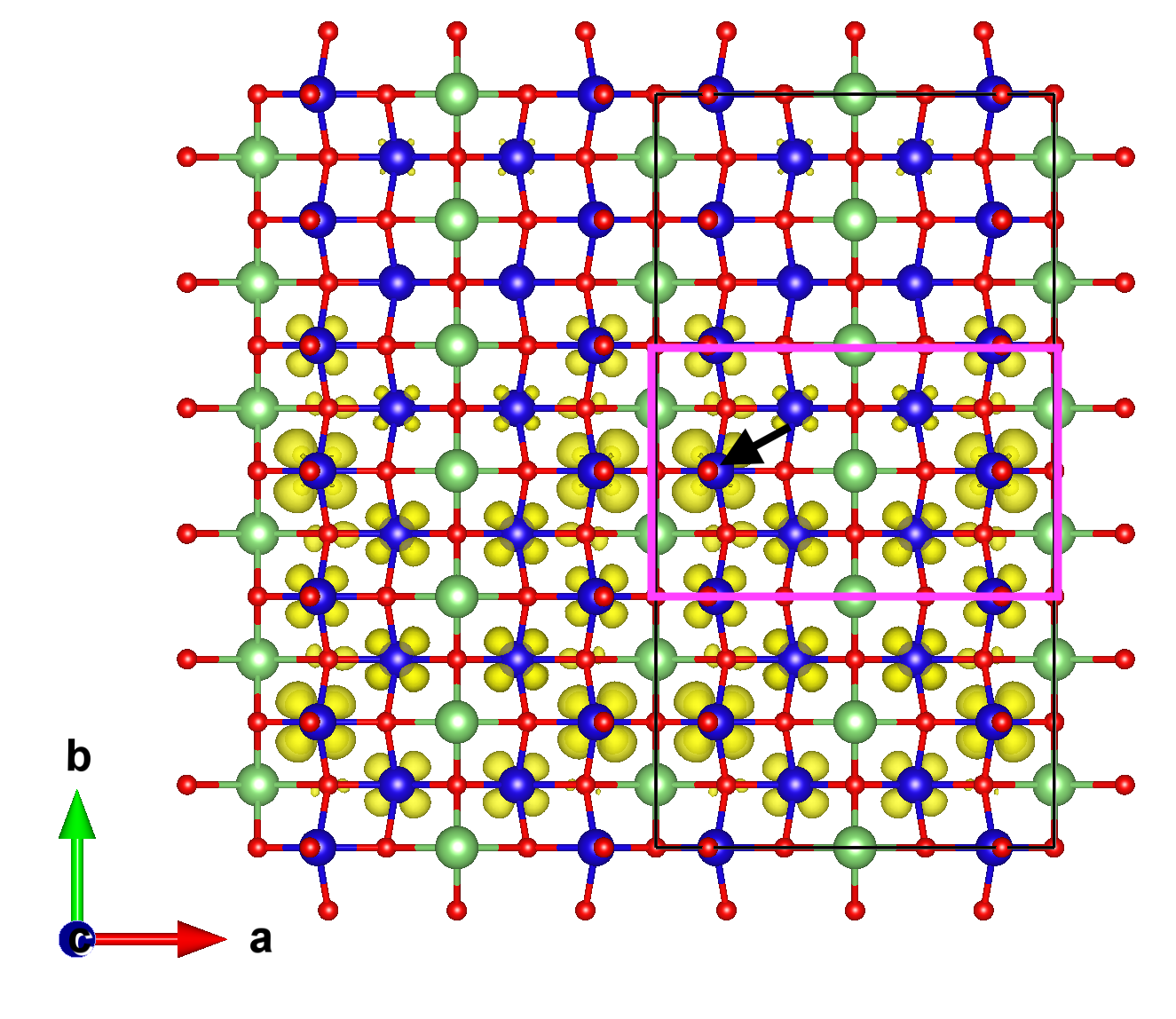}
  \caption{Probability density of holes for electron placed on the V indicated by the arrow, for excitonic states in the range 1.7-1.9 eV. \label{fig-expsi}}
  \end{figure} 
  
Next, we consider the optical properties calculated in BSE and the independent particle approximation (IPA) in Fig. \ref{figopticsafm} compared to those of pure V$_2$O$_5$. 
We note that for  ${\bf E}\parallel{\bf a}$ there is a very strong absorption band near 2 eV, which is already prominent in the IPA but as usual red-shifted and modified in shape with two peaks in the BSE. To allow for comparison with pure V$_2$O$_5$ we show it on a different  vertical scale in the middle panel. This allows us to better see the energy range above 3 eV. We may note that apart from a upward shift of about 0.6 eV the spectrum in this range is similar to that of pure V$_2$O$_5$. This shift is related to the flattening of the VBM in LiV$_2$O$_5$  noted earlier for the
band structures. For ${\bf E}\parallel{\bf b}$, the spectrum shows   a much smaller absorption near 2 eV and at higher energies the spectrum resembles that of  pure V$_2$O$_5$.

In Fig \ref{figopticsfm} we show the optical dielectric function for the FM case. We see a similar strong absorption band near 2 eV for the 
${\bf E}\parallel{\bf a}$ polarization.  These calculations include 32 valence bands (VB) and 20 conduction bands (CB) for the FM cell, 64 VB and 40 CB  for the AFM cell, which contains twice the number of atoms,  and 30 VB, 20 CB for  V$_2$O$_5$. Including all the O-$2p$ derived and V-$3d$ bands is important for the convergence of these spectra. 

The main new feature, which has its first peak  more precisely at  1.85 eV,  corresponds to the transitions between the occupied split-off band to the lowest unoccupied
conduction band, as can be seen in Fig. \ref{figbndwt}.
 This feature is similar to what was pointed out in NaV$_2$O$_5$ in Bhandari \etal\cite{Bhandari_2015_Na} and was in that material also studied by electron loss spectroscopy\cite{Atzkern2001}. In NaV$_2$O$_5$, experimentally  it lies at about 1 eV rather than near 2 eV. However, the QS$GW$ gap between these states is also  near 2 eV similar to the situation in LiV$_2$O$_5$. Thus, strong excitonic effects again influence this optical transition as we can see by comparing BSE with
 IPA. Closer inspection of the excitonic eigenvalues, shows that there are also nearly dark excitons starting at about 1.4 eV. 

A minor difference may result between Na and Li from the difference in distortion of the structure. 
The angle between the O$_b$-V-$O_v$ gives a measure of the tilting of the pyramid with its O$_v$ towards the inserted alkali cation and is 102$^\circ$ for Li and 104$^\circ$ for Na. The sharper angle means more tilting. The distance from the vanadyl O$_v$ toward the Na is 2.707 \AA\ 
while toward the Li is 2.66 \AA. These stronger distortions are consistent with a slightly stronger downward shift of the
occupied V-$d_{xy}$ like state by interacting with the intercalated cation. 
 
The reason this absorption  is polarized almost exclusively along $a$ is that it corresponds to transitions between odd and even states  with respect to the
mirror-plane passing through  the bridge O and perpendicular to the $a$ direction as was pointed out above and explained in more detail in \cite{Bhandari15,Bhandari_2015_Na} Since this is a transition between the molecular localized type states centered on the bridge O and having mainly V-$d_{xy}$ character for both states,
but with a O-$2p$ antibonding character for the upper empty band, 
   it has a strong intra-atomic (on V) as well as  localized character confined to the rung of the ladder and is therefore  significantly stronger
   than the charge-transfer type transitions between the main O-${2p}$ VBM and V-$3d$ like CBM in pure V$_2$O$_5$.
   This can be seen in Fig.\ref{fig-expsi}. We can clearly see the $xy$ character of the states. The strongest hole contribution is on the V where the electron is placed and its neighbor across the V-O-V rung.

   The very strong peak in $\varepsilon_2(\omega)$ in the $a$-direction, also leads to a strong enhancement of the $\varepsilon_1(0)=46.3$ in the BSE and $42.4$ in the IPA in the AFM. These are the static values but including only electronic screening without contributions from the phonons, commonly referred to as $\varepsilon_\infty$. This compares with a value of 6.2 in the $b$-direction in BSE and $5.4$ in the IPA. In pure V$_2$O$_5$, the electronic screening only dielectric constant were calculated using a different approach in \cite{Bhandari14} and are 6.54 in $a$ and 6.08 in the $b$ directions.  This is a remarkable anisotropy in the dielectric constant.

   \section{Discussion of the cathodoluminscence}
\begin{figure}
  \includegraphics[width=8.75cm]{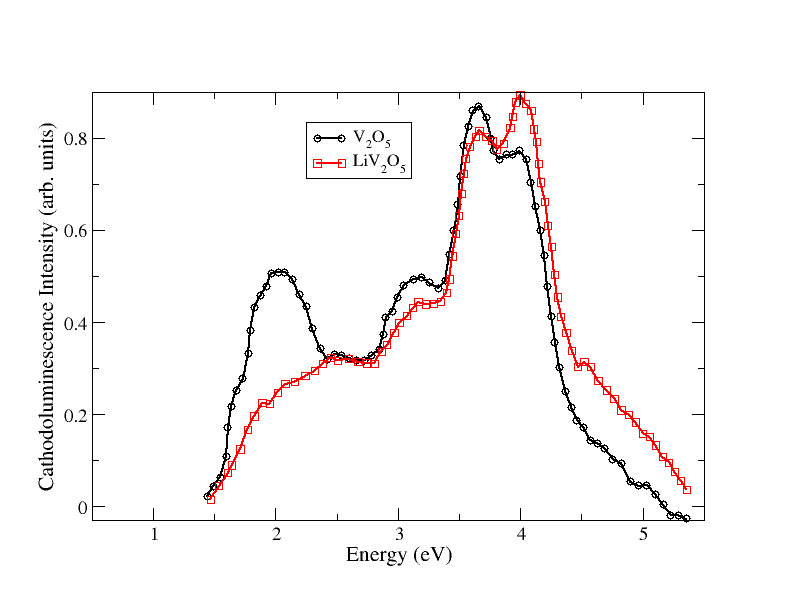}
  \caption{Digitized experimental data of Cathodoluminescence for V$_2$O$_5$
    and LiV$_2$O$_5$ from Walker \etal \cite{Walker_2020}, Fig. 3a and Fig. 3b curves for 8 nm penetration depth.\label{figCL}}
\end{figure}

   While our main results pertain to the absorption, we here wish to comment on the related cathodoluminescene (CL) spectra, which are the main experimental data on optical properties available at this time. 
Walker et al\cite{Walker_2020} studied  changes in the electronic structure of
Li$_x$V$_2$O$_5$ using depth resolved CL. Each of their peaks shows considerable fine structure that evolves with depth from bulk to surface and is also
shown to be sensitive to remote oxygen plasma treatment, indicating that
O-vacancies play a role.  It is here not our intent to interpret these fine
changes but rather to understand the broad main features and their  changes between
pure V$_2$O$_5$ and LiV$_2$O$_5$.  Their spectrum for pure V$_2$O$_5$
, which is here reproduced approximately in Fig. \ref{figCL} by digitizing the data from their figure, shows a
first peak just below 2 eV with fine structure at 1.8 and 1.9 eV, followed by a smaller peak near 3 eV and merging peaks at 3.6 and 4.1  eV.  The main change upon lithiation is that the feature near
2 eV is strongly suppressed and essentially absent but can be partially restored by delithiation.   These authors interpreted
the feature near 2 eV as resulting from recombination of V-$d_{xy}$ like
split-off conduction electrons recombining with holes at the VBM while the higher
peaks were associated with other V-$d$ $t_{2g}$ orbitals, such as $d_{xz}$ and $d_{yz}$.

Considering first Walker \etal's 
interpretation, it is not clear why the observed features are associated with density of states features in the conduction band only while ignoring the valence band structure and assuming that all holes have relaxed  to the VBM. This would
require one to assume a much faster hole than electron relaxation, for which there is no direct evidence, to the best of our knowledge.
Second, the feature  below 2 eV
is about 0.5 below the indirect absorption onset which was found to be near 2.35 eV by
Kenny and Kannewurf\cite{Kenny66}. It is thus clearly not a pure band to band transition or related band gap exciton but may be related to defects. 
Furthermore,  adding  Li would partially
fill the split-off band and  while one could imagine this to suppress
absorption to these states, it would not really affect luminescent recombination between the split-off band and the VBM. The main observation of a decrease in the CL peak below 2 eV thus remains unexplained.  Here we propose an alternative
explanation. 

Unfortunately, the interpretation of CL spectra is rather complex and a fully first-principles approach is not yet available. In a CL experiment electron hole pairs are first created by a high energy electron impact and then the electrons and holes  relax to their  lowest states near the band edges by multiple phonon emissions 
before they radiatively recombine and emit light. That last process is also in competition with non-radiative recombination. The spontaneous luminescence rate $R_e(\omega)$ arising from the perfect crystal energy bands bands is proportional to the corresponding absorption coefficient with $\alpha(\omega)=\varepsilon_2(\omega)\omega/n(\omega)c$ and 
$R_e(\omega)\propto\omega^3\varepsilon_2(\omega)/\left[e^{\hbar\omega/k_BT}-1\right]$. Here, $n(\omega)$ is the real part of the index of refraction and $k_B$ is Boltzmann's constant. However, as the excited electrons relax they increasingly populate lower energy states and hence for a defective crystal CL will more prominently show recombination from the defect levels in the gap to which the excited electrons trickle down. 

The optical absorption in pure V$_2$O$_5$ has recently been shown to be dominated by very strongly bound excitons \cite{Gorelov22,Garcia_2024}. 
The first strong absorption occurs at 3 eV followed by peaks at 4 eV and 5-6 eV with the quasiparticle gap at $\sim3.8$ eV in the QS$G\hat{W}$  method used in \cite{Garcia_2024} and at slightly higher energies $4.4$ eV for the quasiparticle gap and 3.4 eV for the lowest exciton in QS$GW$ as seen in Fig. \ref{figopticsafm}. 
We thus associate the  CL peaks between 3-4 eV with the direct excitons and band gap emission.
The absorption onset of V$_2$O$_5$ near 2.3-2.6 eV on the other hand is
still not entirely understood but is likely associated with indirect
excitons and/or a dark exciton becoming activated by phonons or disorder
as discussed in \cite{Gorelov23,Garcia_2024}.
The CL feature  at 1.8-1.9 eV is clearly well below the onset of absorption. Thus, it  must rather be related to recombination from a  defect level in the  gap. The most prevalent defect levels in V$_2$O$_5$ are related
to oxygen vacancies and self-trapped polarons. In fact both correspond to electrons trapped on a single V. 
We therefore tentatively associate the lowest  CL peak  with a emission from a self-trapped polaron recombining with a hole in the VB, which could alternatively be described as a self-trapped exciton.  In fact, it has been shown by various calculations \cite{Scanlon08,Ngamwongwan_2021,Watthaisong_2019}
that strong electron-phonon coupling leads to the formation of small electron polarons in V$_2$O$_5$. The added electrons are usually associated with O-vacancies, in particular the vanadyl-O-vacancy. Scanlon \etal\cite{Scanlon08}
showed that O-vacancies
produce defect states at about 1 eV above the VBM similar to Li interstitials when a DFT$+U$ calculation is
used which favors the formation of localized states and polaronic deformation. This was confirmed by Ngamwongwan \etal\cite{Ngamwongwan_2021} and Watthaisong \etal\cite{Watthaisong_2019} who showed that self-trapping of excitons
occurs near an added electron even without the presence of O-vacancies or nearby  Li interstitials. In their
calculations, the self-trapped polaron
is found near the middle of the gap at about 1 eV above the VBM,  while the O-vacancy related states depend on which type of O is removed. It is not straightforward to extract
the energy position of this band from the GGA+U calculations of  \cite{Scanlon08,Ngamwongwan_2021,Watthaisong_2019}
as all of these miss the strong increase in quasiparticle gap we find in our QS$GW$ calculations. However, we may
assume that the polaron band lies close to the position of our Li induced occupied band in LiV$_2$O$_5$ since this also corresponds to adding an electron which localizes on a  V-$d_{xy}$ state, although it would in the present case be spread over two V. It is indeed
about 2 eV above the VBM. Localization on a single V would pull it slightly further down. While in our discussion above, this  was associated with the spin splitting of the split-off band, the relaxation of the atoms near the Li which tilt the O-pyramids near each V toward the Li
play also a decisive role in this feature being shifted down toward the middle of the gap. In summary, we associate the $\sim2$  eV
CL band with a transition from this mid gap level to the VBM.  

So, then why is this polaronic emission band reduced upon lithiation? At first one might think that the Li
brings in more electrons and hence more polaron-like states. However, as we showed in the main results of the previous section there is now also a very strong absorption at about the same energy as the polaron emission peak. While in V$_2$O$_5$ even with some defects, the absorption from the defect levels to the conduction band is negligible compared to the emission  resulting from
recombination with the VBM holes, this is no longer the case in LiV$_2$O$_5$, where we found as main result a  very strong optical absorption peak also at $\sim$1.8 eV. In fact, it is about a factor 10 stronger than the lowest exciton peaks in V$_2$O$_5$
and this is due to the more localized character of this optical transition between occupied and empty bands both dominated by the V-$d_{xy}$ orbitals compared to the transition from a V-$d_{xy}$ state to a more delocalized O-$2p$  VB state involved in the emission process.  
We therefore propose that  the emission is reduced by the strong
absorption at this  same energy. 
In other words, we propose that it is the phenomenon of ``self-reabsorption'' that reduces the
cathodoluminescence at 1.8 eV. Interestingly, the self-trapped polaron level lies at about the same energy as the Li induced occupied band and this level lies almost exactly mid-gap so that absorption from this level to the CBM occurs at almost the same energy as recombination with a hole from the VBM. 

\section{Conclusions}
In this paper we showed that Li electron donation in $\alpha$-LiV$_2$O$_5$  leads to a  spin splitting of the split-off conduction band which creates an occupied  band in the middle of the gap.  This then leads to strong optical absorption  between equal spin bands  with intra-atomic V-$d$ character. Essentially transitions occur
between antisymmetric and symmetric linear combinations of the $d_{xy}$ orbitals on the two V across
a V-O-V rung, which form strongly localized molecular type states.  This strong absorption is polarized exclusively
along the $a$ direction.  We propose a different interpretation of the CL experiments of Walker \etal\cite{Walker_2020}
with the main peaks at 3-4 eV being associated with the  charge transfer direct excitons of V$_2$O$_5$
and the 1.8 eV feature as an electron self-trapped polaron emission which is suppressed in LiV$_2$O$_5$ by self-absorption. 

\acknowledgements{This work was supported by the U.S. Department of Energy Basic Energy Sciences (DOE-BES) under Grant No. DE-SC0008933. Calculations made use of the High-Performance Computing Resource in the Core Facility for Advanced Research Computing at Case Western Reserve University and the Ohio Supercomputer Center.}

{\bf Data Availability}
The data that support the findings of this article are openly available\cite{datagithub}.
\bibliography{Bib/lmto,Bib/dft,Bib/gw,Bib/v2o5,Bib/BSE,Bib/liv2o5}
\end{document}